\documentclass[twocolumn,showpacs,preprintnumbers]{revtex4}
\usepackage{graphicx}

\begin{document}

\textbf{Comment on \textquotedblleft Entropy of Classical Systems with Long-Range Interactions
\textquotedblright }

In a recent letter [1], T.M. Rocha Filho and coworkers address the very interesting issue
of the entropic form to be used for  Hamiltonians with long-range interactions.  
In our opinion the letter misses several points 
% quite debated in the recent literature  
which are of fundamental importance.
% for a complete discussion of this problem. 
Moreover it contains several statements which are not
correct as explained below.
% true or not corroborated by any evidence. In this comment we discuss 
%these arguments 
% and severely question  the generality of the conclusions of this letter. 

First of all, contrary to several  statements in [1], 
% we want to stress that
 the generalization of standard  statistics
% proposed by one of us 
[2] is neither {\it``meaningless and may lead to wrong conclusions"} nor {\it ``limited in scope"}. %as stated by the authors. 
Actually, it  has proved to be very useful in a variety of physical  situations and many other applications beyond physics 
%as recently reported for example  in refs 
[3]. Standard  Boltzmann-Gibbs (BG) statistical mechanics is for sure limited to equilibrium situations where all the  possible microstates are equally probable. On the other hand, nonextensive statistics represents one possible generalization for those frequent situations where this is not true.  
%
%Of course there are very many of such possibilities and, although the q-entropy formalism  cannot %be used for all of them, there is plethoric evidence
%that a wide class of phenomena {\it beyond BG microcanonical equiprobability} can surely be %covered.  
%
Moreover, although the form and extensivity of the entropy are addressed in [1],
the authors ignore results [4]
centrally relevant to precisely these questions.  
Finally, the criticisms cited in Refs. [13,14] of [1] have since long been replied in [5].
Coming back 
now to the more specific argument of long-range Hamiltonian systems  discussed in [1], 
%we  point out that 
the authors  do not  focus on the main situations  where nonextensive 
statistics has been applied, i.e. on the {\it nonhomogeneous} quasi-stationary states found in these systems and more specifically on the Hamiltonian Mean Field (HMF) model [6].
In  [1] in fact the homogenity of the quasi-stationary state is always a-priori assumed together with an equiprobability of microstates. 
At variance, the numerical evidence [6-8] 
%and more specifically in refs. [6,7] 
indicates a {\it strong hierarchical  microscopic structure} (very sensitive to the initial conditions) which can also be interpreted within a glassy-like  formalism [7]. In a very recent paper by Morita and Kaneko [9],  a metastable collective oscillation beyond Vlasov analysis has also been observed.  
These  facts, together with  the crucial importance of the neglected finite-size effects,  invalidates the general conclusions of [1].
In [6] it has been clearly shown that  the generalized statistics allows to predict, in a quantitative way and within a coherent frame, the q-exponential decay of the velocity correlation functions and the anomalous diffusion observed.  This is true also for a generalized version of the model [10], 
%where the  range of the interaction can be varied, 
for several system sizes   and many kind of initial conditions. Homogeneous quasi-stationary states, which have an almost exponential velocity correlation function decay, can be also explained  within this same  scenario as a particular case. In general   the prediction $\gamma=2/(3-q)$ for the anomalous diffusion coefficient  has been successfully verified [6].
Although further investigation is   needed,  it is by now certain 
that  the dynamical anomalies observed in the HMF model and long-range Hamiltonian systems go beyond the possible explanation of standard statistical mechanics, and that a new kind of kinetic theory  is needed, as claimed also in [11]. Generalized statistics offers a quite plausible perspective in this direction. Let us finally summarize some of the relevant statements made in [1] that are incorrect for the HMF model: 
(i)  {\it In the limit $N \to \infty$ the interparticle correlations are negligible} is  not always true for long-range interactions and depends on the order of the limits; 
(ii) The supposition that {\it all microstates compatible with the given constraints are equally probable} is invalid, even for the microcanonical ensemble,  for the quasi-stationary state in the presence of long-range interactions for the physically important case where $t \to \infty$ {\it after} $N \to \infty$; strong indications of nonergodicity are available in the literature;
(iii)  {\it The $BG$ entropy is then the correct form to be used} is trivially correct for $N \to \infty$ {\it after} $t \to \infty$, and clearly wrong the other way around, since the distribution of velocities is {\it not} Maxwellian in the quasi-stationary state.

\bigskip\noindent
{\bf A. Rapisarda *, A. Pluchino *  and C. Tsallis **
}

\bigskip\noindent
* Dipartimento di Fisica e  Astronomia and Infn sezione di Catania 
Universit\'a di Catania, Via S. Sofia 64, I-95123 Catania, Italy. \\
\noindent
** Santa Fe Institute, 1399 Hyde Park Road, Santa Fe, NM 87501, USA,   and CBPF,  Rio de Janeiro, Brazil.\\

Electronic addresses:  andrea.rapisarda@ct.infn.it,
alessandro.pluchino@ct.infn.it, tsallis@santafe.edu 

{\small [1] T.M. Rocha Filho et al.,  Phys. Rev. Lett. {\bf 95}, 190601 (2005).}

{\small [2] C. Tsallis,  J. Stat. Phys. {\bf 52}, 479(1988). }

{\small [3] 
J.P. Boon and C. Tsallis eds, {\it Nonextensive Statistical Mechanics:New Trends, New Perspectives }, Europhysics News {\bf 36}, number 6 (2005); see also M. Gell-Mann and C. Tsallis eds, {\it Nonextensive Entropy: Interdisciplinary Applications }, Oxford University Press, New York  (2004).}

 {\small [4] C. Tsallis, M. Gell-Mann and Y. Sato,  Proc. Natl. Acad. Sc. USA {\bf 102}, 15377 (2005); C. Tsallis,  Milan Jour. of Math. {\bf 73}, 145 (2005); L.G. Moyano, C. Tsallis and M. Gell-Mann, Europhys. Lett. (2006), in press, cond-mat/0509229. 
}

 {\small [5] C. Tsallis, Phys. Rev. E {\bf 69}, 038101 (2004) [cond-mat/0304696]; M. Nauenberg, Phys. Rev. E {\bf 69}, 038102 (2004); C. Tsallis, Physica D {\bf 193}, 3 (2004).  
}

{\small [6] A. Rapisarda, A. Pluchino, p. 202 in ref. [3]; A. Pluchino,  A. Rapisarda, Physica A (2006) in press cond-mat/0511570; A. Pluchino, V. Latora, A. Rapisarda, Physica D {\bf 193}, 315 (2004);
V. Latora, A. Rapisarda, C. Tsallis, Phys Rev. E 64, 056134 (2001)}.

{\small [7] A. Pluchino,  V. Latora,  A. Rapisarda,  Phys. Rev. E {\bf 69}, 056113 (2004);  cond-mat/0506665 and cond-mat/0509031.}

{\small [8]  F. Tamarit  and C. Anteneodo, p. 194 in ref. [3];
F.D. Nobre and C. Tsallis,  Phys. Rev. E {\bf 68}, 036115  (2003); M. Sagakami and  A. Taruya, Cont. Mech. and Therm. {\bf 16},279 (2004) and Mon. Not. R. Astron. Soc. {\bf 364}, 990 (2005).}

{\small [9]   H. Morita, K. Kaneko,  Phys. Rev. Lett. {\bf 96}, 050602 (2006).}

{\small [10] C. Anteneodo, C. Tsallis Phys. Rev. Lett. {\bf 80},  5313  (1998); A. Campa, A. Giansanti, D. Moroni, 
J. Phys. Math. Gen. {\bf 36}, 6897 (2003). }

{\small [11]   P-H. Chavanis, Eur. Phys. J. (2006) in press cond-mat/0509726 and references therein.

%------------------------------------------------------------------------------------

\end{document}